\documentstyle[prl,aps,epsf]{revtex}

\begin{document}
\def \beq {\begin{equation}}
\def \eeq {\end{equation}}
\def \bes {\begin{eqnarray}}
\def \ees {\end{eqnarray}}
\def \ni {\noindent}
\def \nn {\nonumber}
\def \z {\tilde{z}}

\title{New features of the thermal Casimir force at
small separations}

\author{
F.~Chen,${}^1$
G.~L.~Klimchitskaya,${}^{2,}$\footnote{On leave from
North-West Polytechnical University,
 St.Petersburg, Russia. 
E-mail: galina@fisica.ufpb.br}
U.~Mohideen,${}^{1,}$\footnote{
E-mail: umar.mohideen@ucr.edu}
and V.~M.~Mostepanenko${}^{2,}$\footnote{On leave from
Research and Innovation Enterprise  ``Modus'',
Moscow, Russia. E-mail: mostep@fisica.ufpb.br}}

\address
{${}^1$Department of Physics, University of California,
Riverside, California 92521\\
${}^2$Departamento de F\'{\i}sica, Universidade
Federal da Para\'{\i}ba,
C.P.~5008, CEP 58059-970,
Jo\~{a}o Pessoa, Pb-Brazil
}
\maketitle

{\abstract{The difference of the thermal Casimir forces at different
temperatures between real metals is shown to increase with a decrease
of the separation distance. This opens new opportunities for the
demonstration of the thermal dependence of the Casimir force.
Both configurations of two parallel plates and a sphere above a plate
are considered. Different approaches to the theoretical description
of the thermal Casimir force are shown to lead to different
measurable predictions.
}}

%\pacs{12.20.Ds, 42.50.Lc, 05.70.-a}
%\large
%\narrowtext

The Casimir effect \cite{1} which is a rare, direct manifestation of the
zero-point oscillations of electromagnetic field has attracted a great
deal of interest in the last few years. This is in part due to
the resurgence of interest in the fundamental
problems connected with the physical vacuum.
In the simplest situation, the Casimir effect is the attractive force
arising between two parallel uncharged metallic plates placed in
vacuum at zero
temperature. This force is unique as it depends on only the fundamental
constants $\hbar$ and $c$, and on the separation between
the plates. No charges or other
interaction constants are involved, which are usually present with
other forces.
The Casimir force arises from the difference of the
zero-point oscillation spectrum in the absence and in the presence of
plates (extensive literature on the subject can be found in
monographs \cite{2,3,4} and in reviews \cite{5,6}).

After the first, order of magnitude, report of
observation of the Casimir
force \cite{7}, a lot of precision measurements have been performed
\cite{8,9,10,11,12,14,15,16}. As a result, the Casimir effect has
acquired much broader impact, being successfully used in nanotechnological
applications \cite{16,17} and for constraining of hypothetical forces
predicted by the Kaluza-Klein supergravity and other promising extensions
to the standard model \cite{18,19,20}. Theoretically, a lot of different
boundaries were considered and the Casimir force between realistic materials
was calculated including effects of surface roughness, finite conductivity
of the boundary metal and nonzero temperature (see \cite{6} for a review).
Note, that although the surface roughness and finite conductivity
corrections have
been already demonstrated experimentally (see \cite{9,10,11,12,15}),
the temperature effect on the Casimir force is yet to be measured.
Even the theoretical treatment of the temperature effect of the
Casimir force is fraught with serious problems in the case of
real metals \cite{21}.

It is common knowledge that at $T=300\,$K the relative thermal 
correction to the Casimir force
(ratio of the thermal correction to the result at $T=0\,$K)  
is an increasing function of the separation distance
and achieves substantial
values only at sufficiently large separations 
(for the configuration of two plates made of ideal metal it contributes 0.16\%
and 2.5\% of the result at $T=0\,$K at separations 1$\,\mu$m and
2$\,\mu$m, respectively; for the configuration of a plate and a sphere the
respective contributions are 2.7\% and 18.2\% \cite{21}). With the increase
of a separation, however, the total force decreases so rapidly that 
thermal corrections to the Casimir force have not yet been measured.

In the present paper we consider the more sensitive ``difference force
measurements'' where the difference in the thermal Casimir
forces  $\Delta F$ at two
different temperatures rather than the absolute value of the
thermal Casimir force is measured. As is proved below for the case of real
metals,
this difference of the thermal Casimir forces (in contrast with the relative
thermal correction) does not increase but decreases with increasing
separation distance. This opens up new opportunities for the
observation of the thermal effect on the Casimir force at the relatively
small separations of about 0.5$\,\mu$m where the 
relative thermal correction itself
is negligible. Both configurations of two parallel plates and a sphere
(spherical lens) above a plate are considered. In conclusion a
realistic experiment is proposed which could help resolve differences
between the various contradictory
theoretical approaches to the description of the thermal
Casimir force between real metals.

We consider first the configuration of two parallel plates made of real
metals
at a separation
distance $a$ in thermal equilibrium at a temperature
$T_1$. For separations $\lambda_p\leq a\leq 2\,\mu$m (where
$\lambda_p$ is the plasma wavelength) and for not too high temperatures
($\leq 350\,$K) the characteristic frequency $c/(2a)$ of the Casimir effect
between good metals, $Au$ for instance, belongs to the region of infrared
optics.
The first Matsubara frequency
$2\pi k_B T/\hbar$, characterizing the thermal effect, also belongs to the
same frequency region ($k_B$ is the Boltzmann constant).
For such frequencies the plasma model for the
dielectric function is applicable
\beq
\varepsilon(\omega)=1-\frac{\omega_p^2}{\omega^2},
\label{1}
\eeq
\ni
where $\omega_p=2\pi c/\lambda_p$ is the plasma frequency.

The thermal Casimir force was found \cite{23,24} by the substitution
of Eq.~(\ref{1}) into the Lifshitz formula \cite{25} written in the form
of a discrete sum over the Matsubara frequencies. For our purposes, the
perturbation result obtained in \cite{24} (see also \cite{21}) is the most
convenient. It is given by
\beq
F_{pp}(a,T)=F_{pp}^{(0)}(a)\left\{
\vphantom{\sum\limits_{i=2}^{6}c_i\frac{\delta^i}{a^i}}
1+\frac{1}{3}\left(\frac{T}{T_{eff}}\right)^4
-\frac{16}{3}\frac{\delta}{a}\left[1-\frac{45\zeta(3)}{8\pi^3}
\left(\frac{T}{T_{eff}}\right)^3\right]
+\sum\limits_{i=2}^{6}c_i\frac{\delta^i}{a^i}\right\},
\label{2}
\eeq
\ni
where $k_B T_{eff}\equiv\hbar c/(2a)$, $\delta\equiv\lambda_p/(2\pi)$,
$F_{pp}^{(0)}(a)=-\pi^2\hbar c/(240a^4)$, $\zeta(z)$ is zeta function,
and coefficients $c_i\>(2\leq i\leq 6)$ are explicitly calculated in
\cite{24} (their exact values are not needed for our present purposes).
In fact Eq.~(\ref{2}) is the perturbation expansion in powers of two
parameters $\delta/a$, where $\delta$ is the skin depth
of electromagnetic oscillations in the metal, and $T/T_{eff}$,
which are small in the above separation range.
One of
these parameters takes into account the finite conductivity of a metal
and the other one the nonzero temperature. It should be noted that there are
no thermal corrections up to $(T/T_{eff})^4$ in the higher order
conductivity
correction terms from the second up to the sixth order.
If one would wish consider $a<\lambda_p$, our parameter $\delta/a$
is not small and it is necessary to use the optical tabulated data
for the complex refractive index to compute the thermal Casimir force
\cite{6}. At $a>2\,\mu$m the low-temperature asymptotic (2) is not
applicable and numerical computations of Ref.~\cite{24} should be
used. The regions $a<\lambda_p$ and $a>2\,\mu$m are not of our
interest here as the first one is not achievable experimentally
for the test bodies of $>1\,$mm size, and within the second the total
Casimir force is too small.

Now let us suppose that the equilibrium temperature is rapidly
changed from $T_1$ to a new temperature $T_2$ such that
$T_2>T_1$. The subject of our interest is the difference of the two
thermal Casimir forces
\beq
\Delta F_{pp}\equiv \Delta F_{pp}(a,T_1,T_2)=F_{pp}(a,T_2)-
F_{pp}(a,T_1).
\label{3}\eeq

Substituting Eq.~(\ref{2}) into Eq.~(\ref{3}), we arrive to
\beq
\Delta F_{pp}=-\Delta^{\! (1)} F_{pp}(T_1,T_2)
\Delta^{\! (2)} F_{pp}(a,T_1,T_2),
\label{4}
\eeq
where
\bes
&&
\Delta^{\! (1)} F_{pp}(T_1,T_2)=
\frac{\pi^2k_B^4(T_2^4-T_1^4)}{45\hbar^3c^3},
\label{5}\\
&&
\Delta^{\! (2)} F_{pp}(a,T_1,T_2)=1+\frac{90\zeta(3)}{\pi^3}
\frac{\delta}{a}\,\frac{T_{eff}}{T_1+T_2}
\left(1+\frac{T_1T_2}{T_1^2+T_2^2}\right).
\nn
\ees
\ni
It is clearly seen that the quantity (\ref{4}) is negative (i.e. has the
same sign as an attractive Casimir force) because with an increase of
temperature the magnitude of the force increases. It should be particularly
emphasized that the magnitude of
$\Delta F_{pp}$ decreases with an increase of separation distance.
This leads to the conclusion that in difference
force measurements the thermal effect of the Casimir force can be
measured more likely at small separations than at large ones where the
relative thermal correction inself is greater.

%\frac{T_1^2+T_1T_2+T_2^2}{T_1^2+T_2^2}

In Fig.~1 the dependence of $\Delta F_{pp}$ on separation distance is
plotted for $Au$ ($\lambda_p=136\,$nm), $T_1=300\,$K, and $T_2=350\,$K
(solid line). In the same figure the result for ideal metal (infinite
conductivity) is shown
as a dashed line. It is seen that at the smallest separation where the
above computations are applicable ($a=0.15\,\mu$m) the difference thermal
effect is more than 9 times stronger than at the largest separation
($a=2\,\mu$m). The low-temperature result for the ideal metal is obtained
from Eqs.~(\ref{4}), (\ref{5}) by putting $\delta=0$. It does not depend on
the separation distance.

Now we consider the configuration of a sphere (spherical lens) above a
plate.
This configuration was used in the most precise experiments on measuring
the Casimir force by means of an atomic force microscope \cite{9,10,11,12}.
Here the free energy is given by the Lifshitz formula. The Casimir force
acting between a plate and a lens can be obtained by means of the so called
proximity force theorem. The relative error introduced by this theorem
is of order $a/R$ \cite{26}, where $R$ is a sphere radius, i.e. it is a
fraction of a percent for separations under consideration, given the large
spheres with $R\sim 1\,$mm to be used in experiment. The perturbation result
obtained in analogy with Eq.~(\ref{2}) is \cite{21,24}
\beq
F_{ps}(a,T)=F_{ps}^{(0)}(a)\left\{
\vphantom{\sum\limits_{i=2}^{6}c_i\frac{\delta^i}{a^i}}
1+\frac{45\zeta(3)}{\pi^3}\left(\frac{T}{T_{eff}}\right)^3
-\left(\frac{T}{T_{eff}}\right)^4
-4\frac{\delta}{a}\left[1-\frac{45\zeta(3)}{2\pi^3}
\left(\frac{T}{T_{eff}}\right)^3+
\left(\frac{T}{T_{eff}}\right)^4\right]
+\sum\limits_{i=2}^{6}{\tilde{c}}_i\frac{\delta^i}{a^i}\right\},
\label{6}
\eeq
\ni
where $F_{ps}^{(0)}=-\pi^3\hbar cR/(360a^3)$, and
${\tilde{c}}_i=3c_i/(3+i)$.

For the two thermal Casimir forces at temperatures $T_1$ and $T_2$ for the
configuration of a sphere above a plate we consider the difference
quantity $\Delta F_{ps}$
defined as in Eq.~(\ref{3}). By the use of Eq.~(\ref{6})
the following result is obtained
\beq
\Delta F_{ps}=-R\Delta^{\! (1)} F_{ps}(T_1,T_2)
\Delta^{\! (2)} F_{ps}(a,T_1,T_2),
\label{7}
\eeq
where
\bes
&&
\Delta^{\! (1)} F_{ps}(T_1,T_2)=
\frac{\zeta(3)k_B^3}{\hbar^2c^2}(T_2-T_1)(T_1^2+T_2^2),
\nn\\
&&
\Delta^{\! (2)} F_{ps}(a,T_1,T_2)=\left(1+
\frac{T_1T_2}{T_1^2+T_2^2}\right)\,\left(1+2\frac{\delta}{a}\right)
-\frac{\pi^3}{45\zeta(3)}\frac{T_1+T_2}{T_{eff}}\left(1+
4\frac{\delta}{a}\right).
\label{8}
\ees

In analogy with the case of two parallel plates, it is seen that at low
temperatures ($T_1,\,T_2\ll T_{eff}$), where the present theory is
applicable,
$\Delta F_{ps}$ is negative as expected. It should again be noted that the
magnitude of $\Delta F_{ps}$ is a decreasing function of the separation
distance. Thus the difference force measurements of the
thermal effect on the Casimir force can be done at small separations rather
than
at large separations.

In Fig.~2 the dependence of $\Delta F_{ps}/R$ versus separation distance is
plotted for $Au$ with $T_1=300\,$K, and $T_2=350\,$K. The case for the
ideal metal with infinite conductivity is shown as a dashed line.
At the smallest separation $a=0.15\,\mu$m the difference thermal effect is
more than 2 times stronger than at $a=2\,\mu$m. In contrast to the case
of two parallel plates, for ideal metals, the quantity $\Delta F_{ps}/R$
decreases with increasing separation distance.

The difference thermal forces considered above are well adapted to resolve
the contradictions between alternative theoretical approaches to the
calculation of the
Casimir force at nonzero temperature.
In the approach used here (see also \cite{21,23,24,27}) the plasma
dielectric
function, valid at the characteristic frequencies $c/(2a)$ and
$2\pi k_BT/\hbar$, is extended for all frequencies. In particular, it was
used to calculate the zero-frequency term of the Lifshitz formula for the
force and free energy.
The alternative approach \cite{28,29} uses the physically correct behavior
$\varepsilon\sim\omega^{-1}$ at small frequencies at nonzero temperature.
As a result, the contribution of the zero-frequency term in the two 
approaches is different and this has given rise to extensive discussion
\cite{30,31,32,33,34,35} in the recent literature.
In Ref.~\cite{34} it was shown that the dependence 
$\varepsilon\sim\omega^{-1}$ leads to surprises such as 
negative values of entropy 
of the fluctuating electromagnetic field between the plates 
and the violation of the Nernst heat theorem.
To avoid this problem, it was suggested in Ref.~\cite{34}
that for the surface separations of order 1$\,\mu$m the behavior
of $\varepsilon$ at characteristic frequencies should be extended for all
frequencies, including the zero frequency. 
This is in fact embedded in the use of the plasma model,
as above, for the dielectric function \cite{23,24}. 
It must be also emphasized
that the physical results obtained in this manner coincide with those
obtained
by means of the Leontovich surface impedance \cite{35}. 

To contrast the behavior of the difference thermal Casimir force
between the two theoretical
approaches, let us consider the experimentally preferred
configuration of a sphere (spherical lens) above a plate.
The difference
of the thermal Casimir forces calculated in the approach of
Refs.~\cite{21,23,24,27} is given by Eqs.~(\ref{7}), (\ref{8}). We now fix
a value of separation, say, $a=0.5\,\mu$m, fix $T_1=300\,$K, and
consider $\Delta F_{ps}$ as a function of $T_2$, where
$T_1\leq T_2\leq 350\,$K.
In the approach of Refs.~\cite{28,29}, the perpendicular polarization does
not contribute to the zero-frequency term of the Lifshitz formula
whereas the contribution of the parallel polarization is the same as the
one given by the plasma model. All other terms of the Lifshitz
formula are the same in both approaches in the low temperature
limit under consideration here. The zero-frequency contribution of the
perpendicular polarized modes into the Casimir force $F_{ps}(a,T)$
calculated in the framework of the Lifshitz formula and plasma model is
given by \cite{21}
\beq
\frac{k_BTR}{8a^2}\int_{0}^{\infty}\!ydy\ln\left[
1-\left(\frac{y-\sqrt{\omega_p^2+y^2}}{y+\sqrt{\omega_p^2+y^2}}\right)^2
e^{-y}\right]
\approx
-\frac{k_BT\zeta(3)R}{8a^2}\left(1-4\frac{\delta}{a}+
12\frac{\delta^2}{a^2}\right)
\label{9}
\eeq
\ni
(higher order terms do not contribute at separations $a\geq 0.5\,\mu$m).
Then, the low-temperature thermal force in the approach of
Refs.~\cite{28,29} is obtained by the subtraction of Eq.~(\ref{9}) from
Eq.~(\ref{6}). As a result, the difference force in the approach of
\cite{28,29} is given by
\beq
\Delta F_{ps}=-R\Delta^{\! (1)} F_{ps}(T_1,T_2)
\Delta^{\! (2)} F_{ps}(a,T_1,T_2)
+\frac{k_B\zeta(3)R}{8a^2}(T_2-T_1)\left(1-4\frac{\delta}{a}+
12\frac{\delta^2}{a^2}\right),
\label{10}
\eeq
where $\Delta^{\! (1)} F_{ps}$ and $\Delta^{\! (2)} F_{ps}$ are defined
in Eq.~(\ref{8}).

In Fig.~3, the quantity $\Delta F_{ps}/R$ is plotted versus $T_2$ at
a separation $a=0.5\,\mu$m and $T_1=300\,$K using the approach of
\cite{21,23,24,27} (solid line) and using the alternative approach of
\cite{28,29}
(dotted line). The result for ideal metal boundaries
practically coincides with the
solid line for the scale used in the figure.
Three significant differences emerge for the $\vert\Delta F_{ps}/R\vert$
as a function of $T_2$ for the two approaches considered.
First, it is seen in Fig.~3 that at $T_2=350\,$K, the value of
$\vert\Delta F_{ps}/R\vert$, given by the dotted line, is more than
6 times larger than that given by the solid line, a difference which
should be measurable
experimentally. Second, it should be
noted that in the approach of \cite{28,29} the
sign of $\Delta F_{ps}$ is positive, i.e. the magnitude of
the thermal Casimir force
decreases with an increase of $T$. Third, the quantity
$\vert\Delta F_{ps}/R\vert$, given by the dotted line, changes rapidly 
with the change in $T_2$.

In conclusion we would like to note that the changes of
the force amplitude predicted above
are of order $10^{-13}\,$N for the sphere of radius
$R=2\,$mm and $T_2-T_1=50\,$K. This difference of temperature can be
achieved by the illumination of the sphere and plate surfaces with
laser pulses. If laser pulse durations of $10^{-2}\,$s are chosen,
calculations show that equilibrium temperatures of $T_1$ and $T_2$
can be achieved for sufficient duration allowing
Casimir force measurements by means of an
atomic force microscope (note that force oscillations of order $10^{-13}\,$N
were demonstrated with a relative error of about 20\% at a 95\%
confidence level
in the recent measurement of the lateral Casimir
force \cite{15}). The proposed experiment having the same accuracy
would also help in resolving the differences
in the alternative theoretical
approaches to the description of the thermal Casimir force between real
metals.

This work was supported by the National Science Foundation and National
Institute for Standards and Technology. G.L.K. and V.M.M. were also
supported by CNPq (Brazil).

%%%%%%%%%%%%%%%%%%%%%%%%%%%%%%%%%%%%%%%

\widetext
\newpage
\begin{figure}[h]
\vspace*{-3cm}
\epsffile{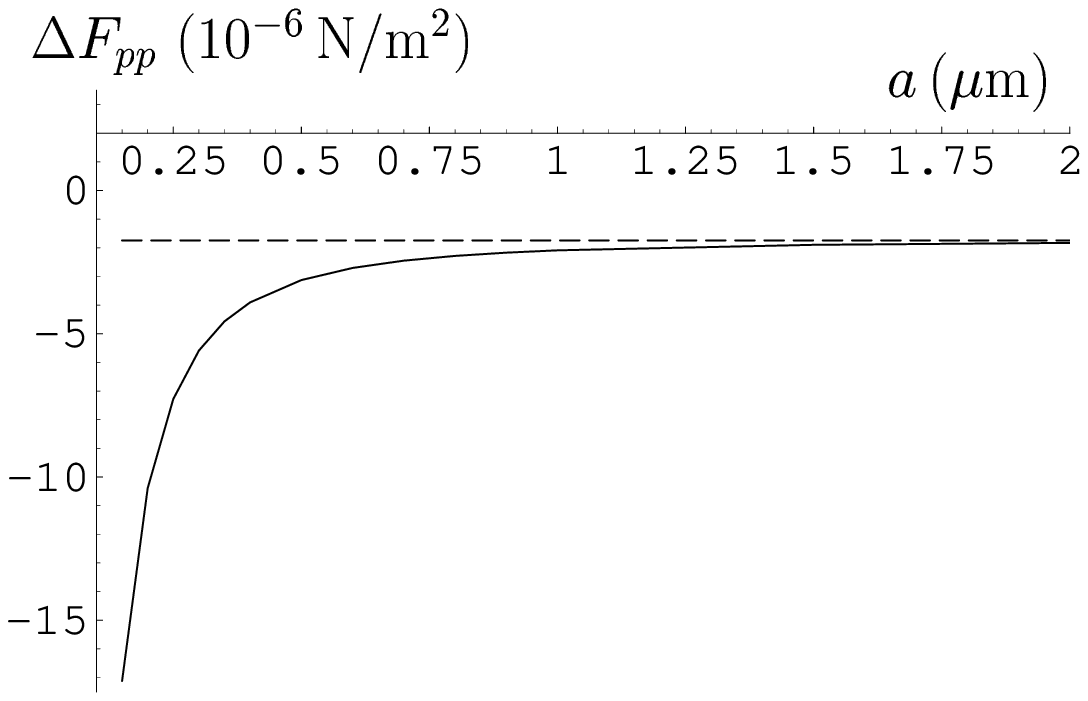}
\vspace*{-8cm}
\caption{Difference of the thermal Casimir forces per unit area
for two parallel plates
versus their separation distance.
The case of the plates made of real metal and that for ideal metal 
 is shown as solid and dashed line, respectively.}
\end{figure}
\newpage
\begin{figure}[h]
\vspace*{-3cm}
\epsffile{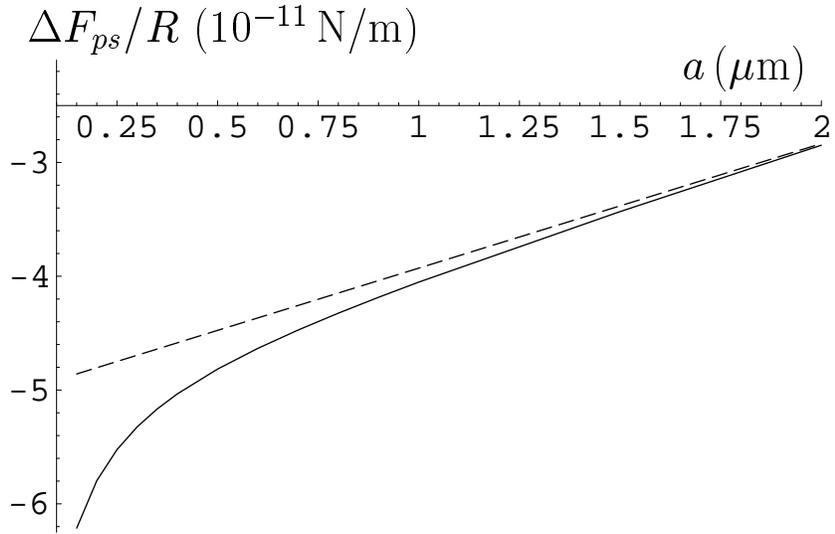}
\vspace*{-8cm}
\caption{Difference of the thermal Casimir forces per unit radius
between a sphere and a plate
versus separation distance.
The case of the bodies made of real metal and that for ideal metal 
 is shown as solid and dashed line, respectively.} 
\end{figure}
\newpage
\begin{figure}[h]
\vspace*{-3cm}
\epsffile{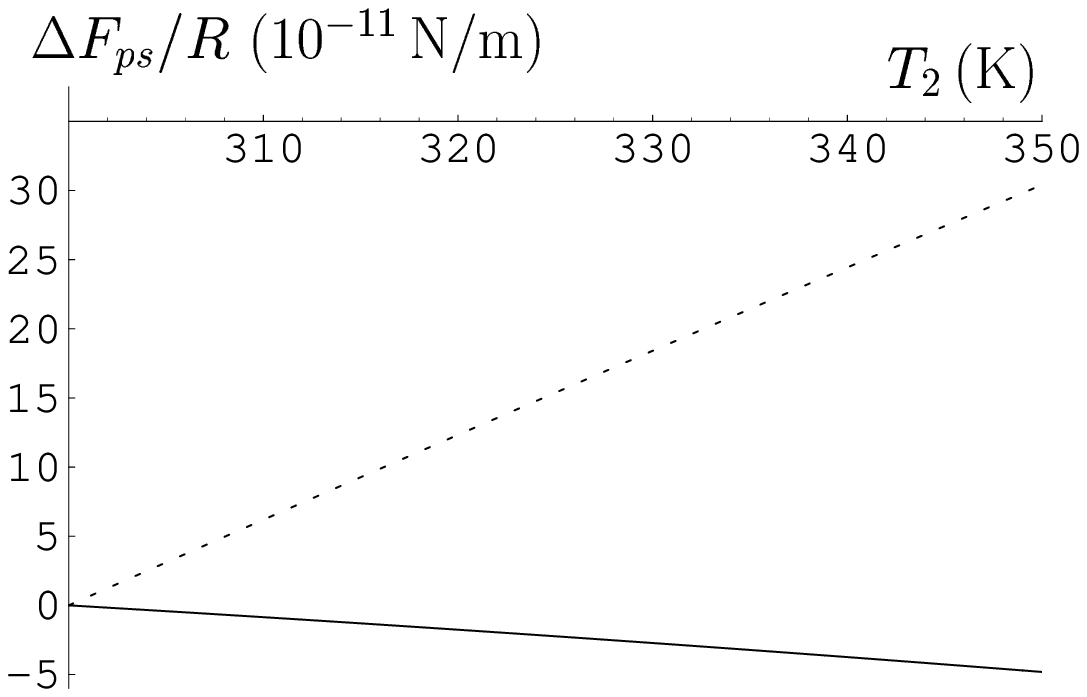}
\vspace*{-8cm}
\caption{Difference of the thermal Casimir forces per unit radius
between a sphere and a plate versus the higher temperature
for the two different theoretical approaches considered.}
\end{figure}


\begin{thebibliography}{99}
\bibitem {1}
H.~B.~G.~Casimir,
{ Proc. K. Ned. Akad. Wet.}
{\bf 51}, 793 (1948).
\bibitem{2}
P.~W.~Milonni,
{\it The Quantum Vacuum}
(Academic Press, San Diego, 1994).
\bibitem{3}
V.~M.~Mostepanenko and N.~N.~Trunov,
{\it The Casimir Effect and its Applications}
(Clarendon, Oxford, 1997).
\bibitem{4}
K.~A.~Milton, {\it The Casimir Effect}
(World Scientific, Singapore, 2001).
\bibitem{5}
M.~Kardar and R.~Golestanian,
Rev. Mod. Phys. {\bf 71}, 1233 (1999).
\bibitem{6}
M.~Bordag, U.~Mohideen, and V.~M.~Mostepanenko,
{ Phys. Rep.} {\bf 353}, 1 (2001).
\bibitem{7}
M.~J.~Sparnaay,
Physica {\bf 24}, 751 (1958).
\bibitem{8}
S.~K.~Lamoreaux,
Phys. Rev. Lett. {\bf 78}, 5 (1997).
\bibitem {9}
U.~Mohideen and A.~Roy,
{ Phys. Rev. Lett.} {\bf 81}, 4549 (1998);
G.~L.~Klimchitskaya, A.~Roy, U.~Mohideen, and V.~M.~Mostepanenko,
{Phys. Rev. A} {\bf 60}, 3487 (1999).
\bibitem {10}
A.~Roy and U.~Mohideen,
{Phys. Rev. Lett.} {\bf 82}, 4380 (1999).
\bibitem {11}
A.~Roy, C.-Y.~Lin, and U.~Mohideen,
{Phys. Rev. D} {\bf 60}, 111101(R) (1999).
\bibitem{12}
B.~W.~Harris, F.~Chen, and U.~Mohideen,
{Phys. Rev. A} {\bf 62}, 052109 (2000).
\bibitem{14}
G.~Bressi, G.~Carugno, R.~Onofrio, and G.~Ruoso,
Phys. Rev. Lett. {\bf 88}, 041804 (2002).
\bibitem{15}
F.~Chen, U.~Mohideen, G.~L.~Klimchitskaya, and
V.\ M.\ Mos\-te\-pa\-nen\-ko,
Phys. Rev. Lett. {\bf 88}, 101801 (2002);
Phys. Rev. A {\bf 66}, 032113 (2002).
\bibitem{16}
H.~B.~Chan, V.~A.~Aksyuk, R.~N.~Kleiman, D.~J.~Bishop, and F.~Capasso,
Science {\bf 291}, 1941 (2001);
Phys. Rev. Lett. {\bf 87}, 211801 (2001).
\bibitem{17}
E.~Buks and M.~L.~Roukes,
{Phys. Rev. B}  {\bf 63}, 033402 (2001).
\bibitem{18}
M.~Bordag, B.~Geyer, G.~L.~Klimchitskaya, and V.~M.~Mostepanenko,
{Phys. Rev. D}  {\bf 58}, 075003 (1998);
{\bf 60}, 055004 (1999);
{\bf 62}, 011701(R) (2000).
\bibitem{19}
V.~M.~Mostepanenko and M.~Novello,
{Phys. Rev. D}{\bf 63}, 115003 (2001).
\bibitem{20}
E.~Fischbach, D.~E.~Krause, V.~M.~Mostepanenko, and M.~Novello,
{Phys. Rev. D}
{\bf 64}, 075010 (2001).
\bibitem{21}
G.~L.~Klimchitskaya
and V.~M.~Mostepanenko,
 { Phys. Rev.} A
{\bf 63}, 062108 (2001).
\bibitem {23}
C.~Genet, A.~Lambrecht, and S.~Reynaud,
Phys. Rev. A {\bf 62}, 012110 (2000).
\bibitem{24}
M.~Bordag, B.~Geyer, G.~L.~Klimchitskaya,
and V.~M.~Mostepanenko,
{ Phys. Rev. Lett.}  {\bf 85}, 503 (2000).
\bibitem{25}
I.~E.~Dzyaloshinskii, E.~M.~Lifshitz, and L.~P.~Pitaevskii,
Sov. Phys. Usp. {\bf 4}, 153 (1961).
\bibitem{26}
M.~Schaden and L.~Spruch,
Phys. Rev. Lett. {\bf 84}, 459 (2000).
\bibitem{27}
B.~Geyer, G.~L.~Klimchitskaya,
and V.~M.~Mostepanenko,
 { Phys. Rev.} A
{\bf 65}, 062109 (2002).
\bibitem{28}
M.~B\"{o}strom and B.~E.~Sernelius,
Phys. Rev. Lett. {\bf 84}, 4757 (2000);
Microelectr. Eng. {\bf 51--52}, 287 (2000).
\bibitem{29}
I.~Brevik, J.~B.~Aarseth, and J.~S.~Hoye,
Phys. Rev. E {\bf 66}, 026119 (2002);
J.~S.~Hoye, I.~Brevik, J.~B.~Aarseth, and K.~A.~Milton,
quant-ph/0212125.
\bibitem{30}
S.~K.~Lamoreaux,
Phys. Rev. Lett. {\bf 87}, 139101 (2001).
\bibitem{31}
B.~E.~Sernelius,
Phys. Rev. Lett. {\bf 87}, 139102 (2001).
\bibitem{32}
B.~E.~Sernelius and M.~B\"{o}strom,
Phys. Rev. Lett. {\bf 87}, 259101 (2001).
\bibitem{33}
M.~Bordag, B.~Geyer, G.~L.~Klimchitskaya,
and V.~M.~Mostepanenko,
Phys. Rev. Lett. {\bf 87}, 259102 (2001).
\bibitem {34}
V.~B.~Bezerra, G.~L.~Klimchitskaya,  and V.~M.~Mostepanenko,
 { Phys. Rev.} A
{\bf 66}, 062112 (2002).
\bibitem {35}
V.~B.~Bezerra, G.~L.~Klimchitskaya, and C.~Romero,
 { Phys. Rev.} A
{\bf 65}, 012111 (2002).

\end{thebibliography}
\end{document}